\documentclass[reprint, amsmath,amssymb, aps,pra]{revtex4-2}
\usepackage{stackengine}
\usepackage{booktabs}
\usepackage{float}

\usepackage[caption=false]{subfig}
\usepackage{graphicx}
\usepackage{bm}
\usepackage{hyperref}
\usepackage{natbib}
\usepackage[usenames,dvipsnames]{color}
\usepackage[usenames,dvipsnames,svgnames,table]{xcolor}
\usepackage{physics}
\usepackage{ulem}

\usepackage{comment}
\bibpunct{\color{blue}[}{\color{blue}]}{,}{n}{}{;}
\hypersetup{
  colorlinks,
  citecolor=blue,
  linkcolor=blue,
  urlcolor=blue
  }

\newlength{\mysize}

\begin{document}
\preprint{APS/123-QED}
\title{Reply to ``Comment on `Trivial Andreev Band Mimicking Topological Bulk Gap Reopening in the Nonlocal Conductance of Long Rashba Nanowires'" }
\author{Richard Hess}
\author{Henry F. Legg}
\author{Daniel Loss}
\author{Jelena Klinovaja}%
\affiliation{%
 Department of Physics, University of Basel, Klingelbergstrasse 82, CH-4056 Basel, Switzerland }%
 \date{\today}
 \begin{abstract} 
 In this Reply we respond to the comment by Das Sarma and Pan \cite{sarma2023} on Hess {\it et al.}, Phys. Rev. Lett. 130, 207001, ``Trivial Andreev Band Mimicking Topological Bulk Gap Reopening in the Nonlocal Conductance of Long Rashba Nanowires" \cite{hess2023}. First, we note that Das Sarma and Pan reproduce the key results of Hess {\it et al.}, substantiating  that our findings are entirely valid. Next, we clarify the incorrect statement by Das Sarma and Pan that the main result of Hess {\it et al.} requires a ``contrived periodic pristine system", pointing out the extensive discussion of positional disorder in the Hess {\it et al.} We also demonstrate that nonlocal conductance features are generically reduced by disorder. This applies to both an Andreev band and to a genuine topological bulk gap reopening signature (BRS). In fact, the suppression of nonlocal conductance of a genuine BRS by disorder was discussed in, {\it e.g.}, Pan, Sau, Das Sarma, PRB 103, 014513 (2021) \cite{pan2021}. We conclude that, contrary to the claims of Das Sarma and Pan, the minimal model of Hess {\it et al.} is relevant to current realistic nanowire devices where only a few overlapping ABSs would be required to mimic a BRS.
\end{abstract}

\maketitle

Das Sarma and Pan  \cite{sarma2023} reproduce the key results from our recent publication Hess {\it et al.}, Phys. Rev. Lett. 130, 207001, ``Trivial Andreev Band Mimicking Topological Bulk Gap Reopening in the Nonlocal Conductance of Long Rashba Nanowires"~\cite{hess2023}. This reproduction demonstrates that overlapping Andreev bound states (ABSs) can result in a nonlocal conductance signal that mimics the bulk gap closing and reopening signal (BRS) expected from the topological superconducting phase transition in Rashba nanowires. 

Despite reproducing our results, Das Sarma and Pan~\cite{sarma2023} claim the findings of Hess {\it et al.}~\cite{hess2023} are ``finetuned [{\it sic}] artificial features" derived from a ``contrived periodic pristine system". This claim is false.  In reality, Hess {\it et al.}~\cite{hess2023} includes extensive discussions about disorder, especially disorder in the ABS position. For example, in the main text of Hess {\it et al.}, it is clearly stated: ``We find the Andreev band in systems with a length of $L\lesssim 10$ localization lengths can tolerate sizeable deviations from a periodic distribution; see the SM for a systematic analysis." Indeed, the Supplemental Material devotes an entire section to disorder in the position of ABSs. As such, the claim of Das Sarma and Pan is unrepresentative of the full findings of Hess {\it et al.} and, furthermore, appears to miss the point of a {\it minimal} model.

In addition, Das Sarma and Pan claim that ``the artificial finetuned [{\it sic}] Andreev band results presented in \cite{hess2023} are suppressed by disorder". Again, this claim ignores the extensive discussions of disorder
in Hess {\it et al.}~\cite{hess2023}. Furthermore, it does not acknowledge that nonlocal features are generically suppressed by  disorder and smooth potentials, including the genuine BRS from a topological phase transition \cite{pan2021,hess2021,hess2023}. Indeed, Pan, Sau, Das Sarma, PRB 103, 014513 (2021) \cite{pan2021}, conclude the following about a genuine BRS from a topological phase transition: ``Disorder generally suppresses the magnitude of the nonlocal conductance everywhere." In fact, in Ref.~\citenum{pan2021}, in order to compare a genuine BRS in a pristine system (Fig.~2 of Ref.~\citenum{pan2021}) and disordered system (Fig.~5 of Ref.~\citenum{pan2021}) it was necessary for the authors to reduce the scale-bar of the nonlocal conductance by three orders of magnitude -- from $0.1\:e^2/h$ to $10^{-4}\:e^2/h$ -- to observe any nonlocal signal.

\begin{figure*}[!]
\includegraphics[width=0.53\textwidth]{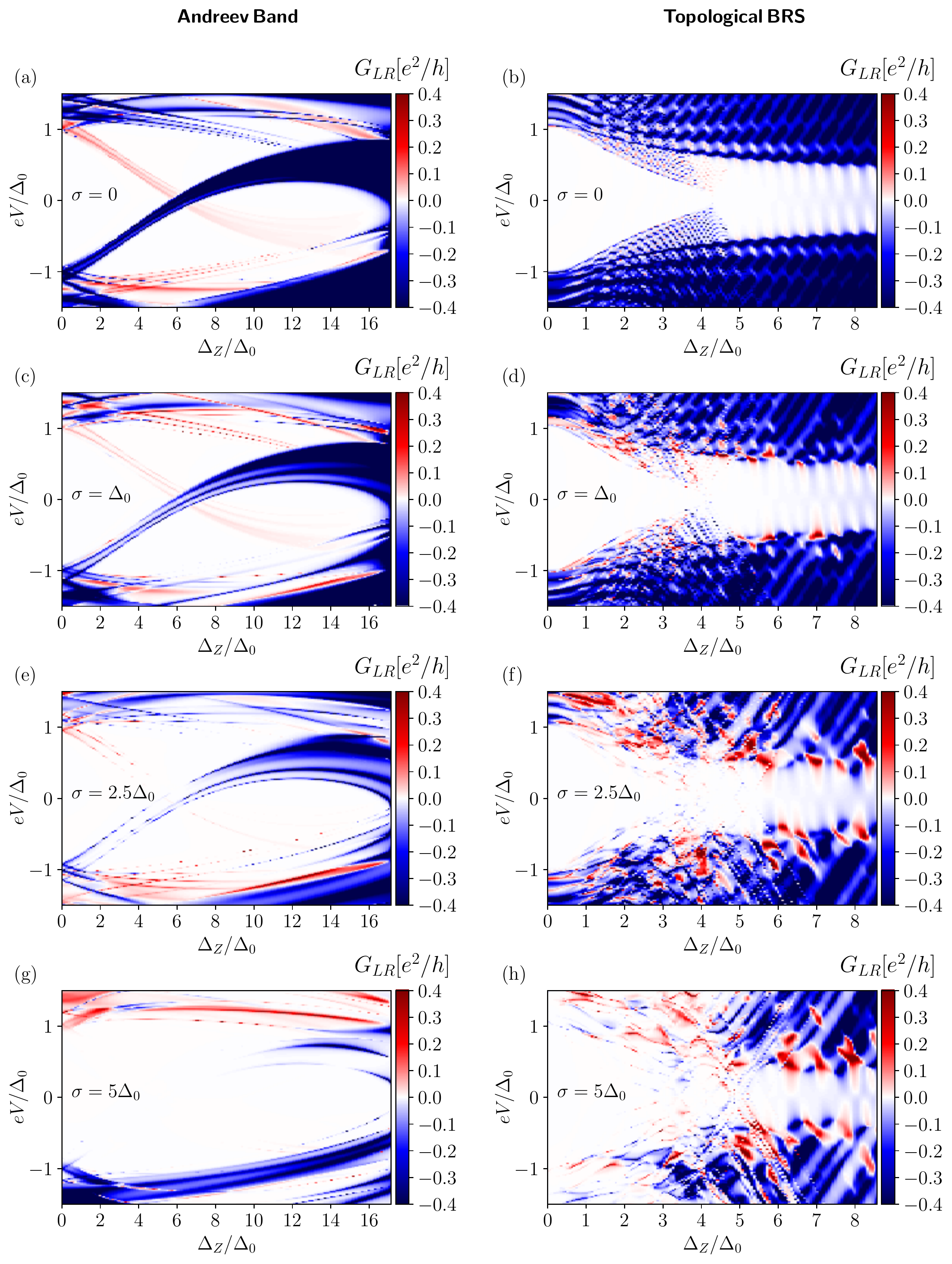}
\caption{{\it Effect of disorder on the nonlocal differential conductance of the Andreev band (left column) and a genuine topological BRS (right column)}. The model used is the same as described  in Hess {\it et al.}~\cite{hess2023}. Disorder is incorporated by additional random onsite potential described by a Gaussian distribution with a standard deviation  $\sigma$. The first row shows the pristine case, while we chose disorder strengths $\sigma= \Delta_0$,  $\sigma= 2.5\Delta_0$, and $\sigma =5 \Delta_0$ for the second, third, and fourth row, respectively. The Andreev band exhibits stability against a reasonable amount of disorder ($\sigma \lesssim 2.5 \Delta_0$) but  disappears for very strong disorder, see the case $\sigma = 5\Delta_0$. The genuine topological BRS also remains visible in nonlocal conductance for small disorder. However, for stronger disorder, the nonlocal conductance signal decreases, especially at small Zeeman fields, and the BRS feature becomes far less clear compared to the pristine case. 
Parameters: lattice constant $a=4$ nm, chemical potential $\mu=2.5$ meV,  superconducting gap $\Delta_0=0.6$ meV, effective mass $m_{\rm{eff}}=0.015 m_e$ with $m_e$ the electron mass. Chemical potential in the lead: $10$ meV. Length of the barriers: $L_b=20$ nm, potential of the barrier $\gamma=5$ meV, temperature $T=0$ K. We chose linear grids of 150 and 1001 data points for the Zeeman field strength and the bias, respectively.
Additional parameters of the trivial Andreev band: Spin orbit interaction strength 
 $\alpha=0$ eV\AA,  length of the normal sections forming the Andreev bound states: $L_N=40$nm, length of the superconducting sections: $L_S=600$nm, number of superconducting sections $M=6$, and we chose the two outer superconducting sections of $L_S/2$. The total length of the nanowire is therefore $L=3200$ nm excluding the barriers. Critical Zeeman field strength $\Delta_Z^c=17.14 \Delta_0$.
Additional parameters of the topological system: Spin-orbit interaction strength 
 $\alpha=0.35$ eV\AA, length of the wire: $L= 3200$ nm  excluding the barriers. We did not use an upper critical field $\Delta_Z^c$ in the topological system.
\vspace{-20pt}
\label{Fig:DisorderAndreevBandVsTopoSys} }
\end{figure*}

To evince how nonlocal conductance features are suppressed by the presence of potential disorder, in Fig.~\ref{Fig:DisorderAndreevBandVsTopoSys}, we present nonlocal conductance for both an Andreev band and genuine BRS for various disorder strengths. As expected, the Andreev band and genuine BRS exhibit some stability against disorder but strong potential disorder suppresses the nonlocal conductance features in both cases.  We also comment that the resolution of the nonlocal conductance grid in bias voltage and magnetic field can strongly affect the clarity of features in conductance color maps.

Finally, we note that in Ref.~\citenum{sarma2023}, the case of ``larger'' disorder has suppressed virtually all nonlocal conductance signal, including from features unrelated to the Andreev band, such that even the superconducting band edge ($|V_{\rm bias}|\approx \Delta $) is no longer visible.

To summarize: 1) The claim~\cite{sarma2023}  that Hess {\it et al.} requires a ``contrived periodic pristine system'' is false and ignores the extensive discussions about disorder in Hess {\it et al.}~\cite{hess2023}, including the entire section on positional disorder in the Supplemental Material. 2) The claim~\cite{sarma2023}  that the Andreev band is suppressed by disorder is generically true of any nonlocal conductance feature. It is therefore also true of a genuine BRS due to a topological phase transition, as previously discussed by Das Sarma and Pan themselves in their own publications.

As such, we wholly reject the assertions of Das Sarma and Pan that our results are ``contrived'',``misleading'', and ``have no implications for any experimental results in actual Majorana nanowires''. In reality, our results show that, in current nanowires that are only a few times longer than the localization length, a few overlapping ABSs can mimic a topological BRS in nonlocal conductance. Our findings therefore raise questions as to the reliability of nonlocal conductance for detecting the phase transition to topological superconductivity.

\section*{Acknowledgements}
\vspace{-10pt}
The authors declare no competing financial or other interests. 
\vspace{-10pt}
\bibliographystyle{apsrev4-1}
\bibliography{commentrefs}

\begin{thebibliography}{4}%
\makeatletter
\providecommand \@ifxundefined [1]{%
 \@ifx{#1\undefined}
}%
\providecommand \@ifnum [1]{%
 \ifnum #1\expandafter \@firstoftwo
 \else \expandafter \@secondoftwo
 \fi
}%
\providecommand \@ifx [1]{%
 \ifx #1\expandafter \@firstoftwo
 \else \expandafter \@secondoftwo
 \fi
}%
\providecommand \natexlab [1]{#1}%
\providecommand \enquote  [1]{``#1''}%
\providecommand \bibnamefont  [1]{#1}%
\providecommand \bibfnamefont [1]{#1}%
\providecommand \citenamefont [1]{#1}%
\providecommand \href@noop [0]{\@secondoftwo}%
\providecommand \href [0]{\begingroup \@sanitize@url \@href}%
\providecommand \@href[1]{\@@startlink{#1}\@@href}%
\providecommand \@@href[1]{\endgroup#1\@@endlink}%
\providecommand \@sanitize@url [0]{\catcode `\\12\catcode `\$12\catcode
  `\&12\catcode `\#12\catcode `\^12\catcode `\_12\catcode `\%12\relax}%
\providecommand \@@startlink[1]{}%
\providecommand \@@endlink[0]{}%
\providecommand \url  [0]{\begingroup\@sanitize@url \@url }%
\providecommand \@url [1]{\endgroup\@href {#1}{\urlprefix }}%
\providecommand \urlprefix  [0]{URL }%
\providecommand \Eprint [0]{\href }%
\providecommand \doibase [0]{http://dx.doi.org/}%
\providecommand \selectlanguage [0]{\@gobble}%
\providecommand \bibinfo  [0]{\@secondoftwo}%
\providecommand \bibfield  [0]{\@secondoftwo}%
\providecommand \translation [1]{[#1]}%
\providecommand \BibitemOpen [0]{}%
\providecommand \bibitemStop [0]{}%
\providecommand \bibitemNoStop [0]{.\EOS\space}%
\providecommand \EOS [0]{\spacefactor3000\relax}%
\providecommand \BibitemShut  [1]{\csname bibitem#1\endcsname}%
\let\auto@bib@innerbib\@empty
\bibitem [{\citenamefont {Sarma}\ and\ \citenamefont {Pan}(2023)}]{sarma2023}%
  \BibitemOpen
  \bibfield  {author} {\bibinfo {author} {\bibfnamefont {S.~D.}\ \bibnamefont
  {Sarma}}\ and\ \bibinfo {author} {\bibfnamefont {H.}~\bibnamefont {Pan}},\
  }\href@noop {} {\  (\bibinfo {year} {2023})},\ \Eprint
  {http://arxiv.org/abs/2306.10041} {arXiv:2306.10041 [cond-mat.mes-hall]}
  \BibitemShut {NoStop}%
\bibitem [{\citenamefont {Hess}\ \emph {et~al.}(2023)\citenamefont {Hess},
  \citenamefont {Legg}, \citenamefont {Loss},\ and\ \citenamefont
  {Klinovaja}}]{hess2023}%
  \BibitemOpen
  \bibfield  {author} {\bibinfo {author} {\bibfnamefont {R.}~\bibnamefont
  {Hess}}, \bibinfo {author} {\bibfnamefont {H.~F.}\ \bibnamefont {Legg}},
  \bibinfo {author} {\bibfnamefont {D.}~\bibnamefont {Loss}}, \ and\ \bibinfo
  {author} {\bibfnamefont {J.}~\bibnamefont {Klinovaja}},\ }\href {\doibase
  10.1103/PhysRevLett.130.207001} {\bibfield  {journal} {\bibinfo  {journal}
  {Phys. Rev. Lett.}\ }\textbf {\bibinfo {volume} {130}},\ \bibinfo {pages}
  {207001} (\bibinfo {year} {2023})}\BibitemShut {NoStop}%
\bibitem [{\citenamefont {Pan}\ \emph {et~al.}(2021)\citenamefont {Pan},
  \citenamefont {Sau},\ and\ \citenamefont {Das~Sarma}}]{pan2021}%
  \BibitemOpen
  \bibfield  {author} {\bibinfo {author} {\bibfnamefont {H.}~\bibnamefont
  {Pan}}, \bibinfo {author} {\bibfnamefont {J.~D.}\ \bibnamefont {Sau}}, \ and\
  \bibinfo {author} {\bibfnamefont {S.}~\bibnamefont {Das~Sarma}},\ }\href
  {\doibase 10.1103/PhysRevB.103.014513} {\bibfield  {journal} {\bibinfo
  {journal} {Phys. Rev. B}\ }\textbf {\bibinfo {volume} {103}},\ \bibinfo
  {pages} {014513} (\bibinfo {year} {2021})}\BibitemShut {NoStop}%
\bibitem [{\citenamefont {Hess}\ \emph {et~al.}(2021)\citenamefont {Hess},
  \citenamefont {Legg}, \citenamefont {Loss},\ and\ \citenamefont
  {Klinovaja}}]{hess2021}%
  \BibitemOpen
  \bibfield  {author} {\bibinfo {author} {\bibfnamefont {R.}~\bibnamefont
  {Hess}}, \bibinfo {author} {\bibfnamefont {H.~F.}\ \bibnamefont {Legg}},
  \bibinfo {author} {\bibfnamefont {D.}~\bibnamefont {Loss}}, \ and\ \bibinfo
  {author} {\bibfnamefont {J.}~\bibnamefont {Klinovaja}},\ }\href {\doibase
  10.1103/PhysRevB.104.075405} {\bibfield  {journal} {\bibinfo  {journal}
  {Phys. Rev. B}\ }\textbf {\bibinfo {volume} {104}},\ \bibinfo {pages}
  {075405} (\bibinfo {year} {2021})}\BibitemShut {NoStop}%
\end{thebibliography}%
\end{document}